# Sustainable AI Regulation

Philipp Hacker[*]

This version: December 21, 2023

Abstract:

This paper addresses a critical gap in the current AI regulatory discourse by focusing on the environmental sustainability of AI and technology, a topic often overlooked both in environmental law and in technology regulation, such as the GDPR or the EU AI Act. Recognizing AI's significant impact on climate change and its substantial water consumption, especially in large generative models like ChatGPT, GPT-4, or Gemini, the paper aims to integrate sustainability considerations into technology regulation, in three steps. First, while current EU environmental law does not directly address these issues, there is potential to reinterpret existing legislation, such as the GDPR, to support sustainability goals. Counterintuitively, the paper argues that this also implies the need to balance individual rights, such as the right to erasure, with collective environmental interests.

Second, cased on an analysis of current law, and the proposed EU AI Act, the article suggests a suite of policy measures to align AI and technology regulation with environmental sustainability. They extend beyond mere transparency mechanisms, such as disclosing GHG footprints, to include a mix of strategies like co-regulation, sustainability by design, restrictions on training data, and consumption caps, potentially integrating AI and technology more broadly into the EU Emissions Trading Regime. Third, this regulatory toolkit could serve as a blueprint for other technologies with high environmental impacts, such as blockchain and Metaverse applications. The aim is to establish a comprehensive framework that addresses the dual fundamental societal transformations of digitisation and climate change mitigation.

**Table of Contents**

---

I. **Introduction**

The next generation of AI systems is currently being developed under the branding of generative AI models (or foundation models).[1] These involve popular applications like ChatGPT, GPT-4, or Midjourney. Training such models is complex and resource intensive.[2] Significantly, not only do they demonstrate the vast transformative potential of AI for society,[3] but they also underscore palpable and emerging risks posed by AI.[4] In this article, the focus will be on a crucial risk dimension that has hitherto been under-appreciated in the legal literature and regulatory debate: AI scholars are increasingly sounding the alarm on the contribution of machine learning to climate change because of its energy[5] and water consumption.[6] For example, recent studies show that the creation of one single image with a leading image generation AI requires as much energy as charging a standard smartphone;[7] by 2027, the total energy consumption of AI is estimated to rival the energy demand of countries such as the Netherlands or Argentina.[8] These results are part of a broader push to map and address the rising contributions of Information and Communication Technologies (ICT) to climate change.[9]

This paper argues that in the wake of these findings AI regulation, and technology law more generally, need to complement its current focus on trustworthiness[10] with an awareness for and the implementation of environmental sustainability in AI and technology development and deployment.[11] Arguably, this is necessary to properly address the dual fundamental societal transitions of digitisation and climate change mitigation.

---

[1] See, e.g., Bommasani and others, "On the opportunities and risks of foundation models", (2021) arXiv preprint arXiv:210807258; Chen and others, "Evaluating large language models trained on code", (2021) arXiv preprint arXiv:210703374. Terminologically, foundation models are trained on large data at scale and are able to tackle diverse tasks; they can be adapted to diverse downstream use cases (ibid, 7). Large AI models overlap mostly–but not fully–with this definition. They typically contain several billion parameters, are trained on large datasets, and require significant compute infrastructure Bienert and others, *Large AI Models for Germany: Feasibility Study* (2023); specifically, generative models output content addressed towards communication, e.g., text, images, videos, or music, see, e.g., Meyer, "ChatGPT: How Does It Work Internally?", *Towards AI* <https://pub.towardsai.net/chatgpt-how-does-it-work-internally-e0b3e23601a1> (last visited 10 December 2022). The terms "large generative AI model" and "foundation model" are used interchangeably in this paper unless specifically differentiated.
[2] Khattab and others, "Demonstrate-Search-Predict: Composing retrieval and language models for knowledge-intensive NLP", (2022) arXiv preprint arXiv:221214024.
[3] See, e.g., for a more positive account, Lobel, *The Equality Machine* (2022).
[4] See, e.g., Kaminski, "Regulating the Risks of AI", 103 BU L Rev (2023), 1347-1411; Hacker, "The European AI Liability Directives - Critique of a Half-Hearted Approach and Lessons for the Future", 51 *Comput. Law Secur. Rev.* (2023), 4-5; Zech, "Risiken Digitaler Systeme", (2020) *Weizenbaum Series* #2; Bubeck and others, "Sparks of artificial general intelligence: Early experiments with GPT-4", (2023) arXiv preprint arXiv:230312712.
[5] See, e.g., Freitag and others, "The real climate and transformative impact of ICT: A critique of estimates, trends, and regulations", 2 *Patterns* (2021), 100340; Schwartz and others, "Green AI", 63 *Communications of the ACM* (2020), 54-63.
[6] Li and others, "Making AI Less 'Thirsty': Uncovering and Addressing the Secret Water Footprint of AI Models", (2023) arXiv preprint arXiv:230403271; Zuccon and others, "Beyond CO2 Emissions: The Overlooked Impact of Water Consumption of Information Retrieval Models", (2023) *Proceedings of the 2023 ACM SIGIR International Conference on Theory of Information Retrieval* 283–289.
[7] Luccioni and others, "Power Hungry Processing: Watts Driving the Cost of AI Deployment?", (2023) arXiv preprint arXiv:231116863, 5 (on Stable Diffusion XL).
[8] de Vries, "The growing energy footprint of artificial intelligence", 7 *Joule* (2023), 2191-2194; Erdenesanaa, "A.I. Could Soon Need as Much Electricity as an Entire Country", *New York Times* (10 October 2023).
[9] See, e.g., Knowles and others, "Our house is on fire: The climate emergency and computing's responsibility", 65 *Communications of the ACM* (2022), 38-40; Freitag and others, op. cit. *supra* note 5; Taddeo and others, "Artificial intelligence and the climate emergency: Opportunities, challenges, and recommendations", 4 *One Earth* (2021), 776-779.
[10] See, e.g., High-Level Expert Group on Artificial Intelligence, *Ethics Guidelines for Trustworthy AI* (2019); Avin and others, "Filling gaps in trustworthy development of AI", 374 *Science* (2021), 1327-1329; Smuha and others, "How the EU can achieve legally trustworthy AI: a response to the European commission's proposal for an artificial intelligence act", (2021) Working Paper, <http://dx.doi.org/10.2139/ssrn.3899991>; Kaur and others,

This paper takes first steps into this direction from a legal perspective. In doing so, it seeks to move beyond the state of the art in three specific respects. First, it analyzes the resources existing environmental and technology regulation may muster to address the climate effects of generative AI models, such as ChatGPT or GPT-4, which have so far received scant attention in the legal literature.[12] The article scrutinizes the European emissions trading and water regulation, before turning to the GDPR and the proposed EU AI Act. The latter has recently entered the final legislative procedure with the trilogue process. Stronger sustainability measures were proposed by the European Parliament (EP) in its position on the Act adopted on June 14, 2023 (AI Act EP Version).[13] The EP, however, chiefly focused on soft rules,[14] such as vague principles,[15] voluntary codes of conduct,[16] green labels,[17] and disclosure rules.[18] These proposals, while laudable, leave significant room for improvement.

Hence, second, the paper develops a comprehensive legal framework to chart AI's, and digital technology's, path towards greater environmental sustainability, in which environmental and technology law mutually learn from each other's regulatory strategies. In legal scholarship, the implications of the law for climate change, and vice versa, are increasingly discussed in the general legal literature.[19] Nonetheless, for technology and AI regulation more specifically, questions of climate change and sustainability still occupy a significant blind spot.[20] In this domain, an interdisciplinary lens is necessary to tease apart the differential impact of machine learning modelling choices for. In this context, I will draw on a growing literature investigating

---

"Trustworthy artificial intelligence: a review", (2022) 55 *ACM Computing Surveys (CSUR)*, 1-38; Laux and others, "Trustworthy Artificial Intelligence and the European Union AI Act: On the Conflation of Trustworthiness and the Acceptability of Risk", (2022) *Regulation and Governance*.

[11] See, e.g., Chien, "Good, better, best: how sustainable should computing be?", (2021) 64 *Communications of the ACM*, 6-7; Güler and Yener, "Sustainable federated learning", (2021) arXiv preprint arXiv:210211274; Perucica and Andjelkovic, "Is the future of AI sustainable? A case study of the European Union", (2022) *Transforming Government: People, Process and Policy*, 347-358; Pagallo and others, "The environmental challenges of AI in EU law: lessons learned from the Artificial Intelligence Act (AIA) with its drawbacks", (2022) *Transforming Government: People, Process and Policy*, 359-376; Hacker, op. cit. *supra* note 4, 63-66.

[12] But see now, e.g., Hacker and others, "Regulating ChatGPT and other Large Generative AI Models", (2023) *ACM Conference on Fairness, Accountability, and Transparency (FAccT '23)*, 1112-1123; Helberger and Diakopoulos, "ChatGPT and the AI Act", (2023) 12 *Internet Policy Review*; Edwards, "Regulating AI in Europe: four problems and four solutions", Ada Lovelace Institute (31 March 2022); Hacker, op. cit. *supra* note 4.

[13] See European Parliament, Report on the proposal for a regulation of the European Parliament and of the Council on laying down harmonised rules on Artificial Intelligence (Artificial Intelligence Act) and amending certain Union Legislative Acts (COM(2021)0206 – C9-0146/2021 – 2021/0106(COD)), Committee on the Internal Market and Consumer Protection and Committee on Civil Liberties, Justice and Home Affairs, Rapporteur: Brando Benifei, Ioan-Dragoş Tudorache, Doc. No. A9-0188/2023, May 22, 2023.

[14] But see the discussion of Art. 9(2)(a) and Art. 28b(2)(a) AI Act EP Version below, Part IVs.2.b).

[15] Art. 4a(2)(h) AI Act EP Version.

[16] Art. 69(2)(g) AI Act EP Version.

[17] Art. 73a AI Act EP Version.

[18] Art. Art. 12(2)(a), 11 and 28b in conjunction with Annex IV(3) AI Act EP Version.

[19] See, e.g., Bosselmann, *The principle of sustainability: transforming law and governance* (Routledge, 2016); Sjåfjell and Richardson (Eds.), *Company Law and Sustainability* (CUP, 2015); Mittwoch, *Nachhaltigkeit und Unternehmensrecht* (Mohr Siebeck, 2022); Zech, "Nachhaltigkeit und Digitalisierung im Recht", 2 ZfDR (2022), 123-134; Schirmer, *Nachhaltiges Privatrecht* (Mohr Siebeck, 2023).

[20] But see Stein, "Artificial intelligence and climate change", (2020) Yale J on Reg, 890-939; Martini and Ruschemeier, "Künstliche Intelligenz als Instrument des Umweltschutzes", (2021) ZUR, 521 et seqq.; Pagallo and others, op. cit. *supra* note 11; see also Hacker, op. cit. *supra* note 4, .

the intersection of computer science, and machine learning more specifically, on the one hand and climate change on the other.[21]

As Dennis Hirsch has famously claimed, data protection law has much to learn from environmental law, particularly concerning the transition from command-and-control to more flexible forms of regulation.[22] This paper seeks to show that this holds as well for AI regulation,[23] and technology regulation more broadly. Conversely, however, environmental law also needs to be updated to cover the climate change risks of emerging technologies. This paper proposes several new policy options necessary to align AI development with sustainability, ranging from co-regulation instruments to sustainability by design, restrictions on AI training, and consumption caps based on an emissions trading system or on an AI model's social utility.

In a third step, the article then uses the toolkit developed for sustainable AI regulation to envision a blueprint for sustainable technology regulation more broadly, for example in the context of blockchain, data centres, or the metaverse.

The remainder of the article is structured as follows. Part II provides an accessible introduction to the benefits and potential risks of AI, with a focus on the environmental dimension. The core topics of the article unfold in Parts III-V. Here, the paper explores AI sustainability under environmental law, the GDPR, and the AI Act (Part III); suggests policy measures for green AI (Part IV); and extends these findings to digital technology more broadly (Part V). The last section concludes (Part VI).

## II. AI and sustainability

While many specific risks concerning AI have been flagged in the past, sustainability considerations have only recently gained traction in the AI research community.

### 1. AI and classical AI risks

In a nutshell, computer scientists continue to disagree on the concept of AI,[24] with one leading definition referring to AI as computer programs that emulate human, rational behaviour and thinking.[25] What is certain, however, is that AI confers tremendous opportunities to our societies,[26] but also harbours several serious risks.[27] Classical risks associated with AI include a range of specific characteristics,[28] such as: data protection and privacy; opacity; discrimination; and unforeseeability. Fifth, large generative AI models in particular may take manipulation, fake news, and hate speech to unprecedented levels by automated mass generation, if not properly reined in.[29] This is another topic now addressed by a growing body of legal research.[30]

---

[21] See, e.g., Bianchini and others, "The environmental effects of the 'twin' green and digital transition in European regions", (2023) 84 *Environmental and Resource Economics*; OECD, "Measuring the Environmental Impacts of AI Compute and Applications", (2022) *OECD Digital Economy Papers*; Cowls and others, "The AI gambit: leveraging artificial intelligence to combat climate change—opportunities, challenges, and recommendations", (2021) 38 *AI & Society*, 283–307; Taddeo and others, op. cit. *supra* note 9; Freitag and others, op. cit. *supra* note 5; Victor Galaz and others, "Artificial intelligence, systemic risks, and sustainability" (2021) 67 *Technology in Society*, 101741; Schwartz and others, "Green AI" (2020) 63 *Communications of the ACM*, 54-63.
[22] Hirsch, "Protecting the inner environment: What privacy regulation can learn from environmental law", (2006) 41 Ga. L. Rev., 1-63.
[23] See also the IPCC risk assessment method suggested for the AI Act in Novelli and others, "Taking AI risks seriously: a new assessment model for the AI Act", (2023) *AI & Society*.
[24] O'Shaughnessy, "One of the Biggest Problems in Regulating AI Is Agreeing on a Definition" *Carnegie Endowment* (6 October 2022).
[25] See, e.g., Russell and Norvig, *Artificial Intelligence: A Modern Approach*, 4th Global ed. (Pearson Education, 2022), pp.19-22.
[26] See refs. in note 3.
[27] See refs. in note 4.
[28] Zech, op. cit. *supra* note 4; Kaminski, op. cit. *supra* note 4; Hacker, "Manipulation by algorithms. Exploring the triangle of unfair commercial practice, data protection, and privacy law" 27 *European Law Journal* (2023), 3-5.

## 2. Another perspective: environmental risks

Beyond theses more traditional risks, however, an increasing body of research points to environmental risks, but also opportunities, posed by AI training and deployment.[31] The current literature operates with a threefold concept of sustainability, branching out into economic, social, and environmental aspects.[32] While these are all important goals, the present paper will focus on the IT law implications of environmental sustainability, which are key to solving the current climate crisis.

Climate change, arguably, poses an existential threat to the human species.[33] Hence, new approaches are needed that go beyond the traditional, but so far largely ineffective international climate summits and individual initiatives of certain climate-progressive states.[34] Effectively, in my view, every legal field–just like every industrial, administrative or consumption sector–will have to chart paths across its own territory to map possible contributions to the collective effort of mitigating climate change.[35] ICT, and the concomitant field of IT law, are particularly well positioned to (co-)lead this effort as ICT arguably has an important role to play concerning both the mitigating and the contributing side of climate change, and IT law is promisingly interdisciplinary in its general approach.

### a) Promises to mitigate global warming

ICT more generally, and AI particularly, may be harnessed to combat climate change in many ways. This is an active field of research in various technical disciplines.[36] It has resulted in numerous theoretical and empirical contributions demonstrating how a reduction of energy, water and material consumption can be achieved by bringing AI applications to bear on questions of project planning, documentation, and implementation.[37] For example, using AI optimisation can significantly reduce the energy consumption necessary for cooling data centres.[38] AI may also power low-carbon energy systems, enhancing energy efficiency and the integration of renewables.[39] This list could easily be extended.

---

[29] See, e.g., the investigation by the European Parliament into the LENSA AI model; see also Oxford Analytica, "Generative AI carries serious online risks" (2023) *Emerald Expert Briefings*.
[30] See, e.g., Hacker and others, op. cit. *supra* note 12; Helberger and Diakopoulos, op. cit. *supra* note 12.
[31] See, e.g., OECD, op. cit. *supra* note 23 et seq. and below, next part.
[32] See, e.g., Giovannoni and Fabietti, "What Is Sustainability? A Review of the Concept and Its Applications" in Busco and others (Eds.), *Integrated Reporting: Concepts and Cases that Redefine Corporate Accountability* (Springer International Publishing, 2013) 28–29.
[33] IPCC, 6th Assessment Report, Climate Change 2022: Mitigation of Climate Change, Summary for Policymakers, C.1.
[34] See Nordhaus, "The climate club: how to fix a failing global effort", 99 *Foreign Affairs* (2020) 10.
[35] Cf. also IPCC, 6th Assessment Report, Climate Change 2022: Mitigation of Climate Change, Summary for Policymakers, C.1–C.11. on the need for an all-encompassing, immediate climate transition in all sectors of society.
[36] See, e.g., Rolnick and others, "Tackling climate change with machine learning" 55 ACM Computing Surveys (CSUR) (2022) 1; Vinuesa and others, "The role of artificial intelligence in achieving the Sustainable Development Goals" 11 *Nature Communications* (2020), 4.
[37] See, e.g., https://www.bcg.com/de-de/publications/2022/how-ai-can-help-climate-change; https://www.sueddeutsche.de/wirtschaft/ki-energie-sparen-industrie-eisengiesserei-aveva-1.5650707; OECD, op. cit. *supra* note 23; Taddeo and others, op. cit. *supra* note 5; Cowls and others, op. cit. *supra* note 23.
[38] DeepMind, "DeepMind AI Reduces Google Data Centre Cooling Bill by 40%" Google DeepMind Blog <https://www.deepmind.com/blog/deepmind-ai-reduces-google-data-centre-cooling-bill-by-40>; Nicola Jones, "How to stop data centres from gobbling up the world's electricity" (2018) 561 *Nature* 163.
[39] Vinuesa and others, op. cit. *supra* note 43, 4.

### b) Contributions of ICT and AI to climate change

On the flip-side, AI and ICT more generally are increasingly recognised as important contributors to climate change in computer science.[40] And with good reason: Current estimates show that ICT contributes up to 3.9% of global greenhouse gas (GHG) emissions[41]–compared to roughly 2.5% for global air travel.[42] The carbon footprint of machine learning more specifically has skyrocketed over the last years.[43] AI training is particularly resource intensive, both in terms of energy and water usage, and even more so with large AI models.[44] However, actually deploying and running a large AI model, such as GPT-4, comes with additional sustainability costs.

It not only consumes energy but also significant amounts of water for cooling the data centres hosting the model: recent estimates posit that a conversation with ChatGPT consisting of 20-50 questions and answers consumes, mostly through evaporation, roughly the content of a 500ml bottle of water, depending on the circumstances of deployment.[45] Given the large number of conversations ChatGPT has powered since its inception, this adds up to a highly significant amount of water–an increasingly scarce resource in many parts of the world (see below, Part III.1.b)).[46] Hence, regulatory intervention and guidance are arguably needed to make AI and technology more environmentally sustainable.

### III. Sustainable AI under current and proposed EU law

The significant climate impact of AI, ICT and technology in general raises the question of the extent to which such costs and risks are contemplated in current and proposed EU law. As we shall see, the consideration of climate costs under current law is at odds with the often-dyadic structure of traditional technology regulation, contemplating each party and its counterparty separately and at a time. While several of regulatory fields do address negative externalities, for example in energy or antitrust law, the bilateral focus is most visible in general contract law: contracts bind the parties, and no one else; effects to third parties are largely ignored.[47] For example, concerning the interpretation of unfair contract terms, the German Federal Court for Private Law (*Bundesgerichtshof*, BGH) ruled that third-party effects may, as a general rule, not be considered in judicial scrutiny of individual clauses.[48] Similarly, the GDPR zooms in on the relationship between data controller and data subject, struggling to address third-party effects of data processing.[49]

---

[40] See n. 9.
[41] Freitag and others, op. cit. *supra* note 5; OECD, op. cit. *supra* note 23, 25-26.
[42] ACM Tech. Policy Council, ACM TechBrief: Computing and Climate Change (2021) 1.
[43] Ibid: potentially by a factor of up to 300,000.
[44] OECD, op. cit. *supra* note 23, 5; Li and others, op. cit. *supra* note 6, 2: "Training GPT-3 in Microsoft's state-of-the-art U.S. data centres can directly consume 700,000 liters of clean freshwater, enough for producing 370 BMW cars or 320 Tesla electric vehicles, and these numbers would have been tripled if GPT-3 were trained in Microsoft's Asian data centres."
[45] Li and others, op. cit. *supra* note 6, 3.
[46] Tzanakakis and others, "Water supply and water scarcity" (2020) 12 *Water* 2347; Dolan and others, "Evaluating the economic impact of water scarcity in a changing world" (2021) 12 *Nature Communications* 1915.
[47] Generally, a third party is any entity affected in a certain situation that is not party to the legal obligation under scrutiny, be it a contractual or a non-contractual obligation. For example, in the context of the GDPR, this concerns any stakeholders except for the data controller and the data subject. The GDPR itself (Article 6(1)(f)) and the CJEU (n. 79) use the term in this sense as well. More specifically, in this paper, the term often denotes persons and entities affected by climate change, but also by the concrete AI application.
[48] BGH, Case VIII ZR 214/80, NJW 1982, 178, 180; more nuanced now BGH, Case III ZR 179/20, NJW 2021, 3179, para. 75: limited consideration of third-party information interests on digital platforms.
[49] See, for the general debate, Ben-Shahar, "Data Pollution" (2019) 11 *Journal of Legal Analysis* 104; Barocas and Levy, "Privacy dependencies" (2020) 95 *Wash L Rev* 555; for the GDPR, Hacker, *Datenprivatrecht: Neue*

This points to the larger challenge of integrating collective rights and interests into a legal system traditionally geared, in market and economic contexts, toward the stabilisation of exchanges between clearly identified parties. To analyse the impact of climate change considerations, the paper, therefore, turns to regulatory fields affecting market exchange, AI development, and environmental costs: environmental law, the GDPR, subjective rights under data protection and non-discrimination law, and the proposed EU AI Act.

### 1. Environmental Law

The first field of law to look for direct regulation of the climate effects of AI and emerging technology is environmental law.[50] The EU's legal regime for environmental protection incorporates an intricate web of regulations, directives, and policy decisions. Of particular note here are the EU Emissions Trading System (ETS[51]) and the Water Framework Directive (WFD[52]).[53] While these norms do not address AI explicitly, they may cover emerging technologies by virtue of their technology-neutral scope and formulation.[54]

#### a) The EU Emissions Trading System

Currently, in the EU, the ETS is the primary tool for reducing emissions, based on a cap-and-trade regime.[55] Compared to a general $CO_2$ tax, for example, the ETS affords the advantage of allowing for the formulation of clear net GHG emissions targets,[56] which seems preferable in times of increasingly urgent decarbonisation.

As the law starts tackling the GHG emissions of AI-related activities, the ETS may be considered a promising vehicle to reduce such emissions. However, such goals are at odds with the ETS' current architecture. At the moment, it targets specific high-consumption sectors, such as power and heat generation, energy-intensive industry sectors (e.g., oil refineries, steelworks, and production of iron, aluminium, cement, paper, and glass), commercial aviation, and maritime transport.[57] Although not explicitly covered, emerging technologies, including AI, can come under the purview of ETS only if operated in sectors regulated by it. This results in indirect limitations on AI utilisation, as exceeding set emission caps will trigger financial sanctions. However, if used outside these sectors, AI and emerging technologies are not covered. Hence, while there is significant potential, the ETS does not currently address the environmental costs of AI and emerging technologies in a direct, comprehensive, and adequate way. This gap will be taken up in the policy proposals to evaluate how the ETS might be reconfigured to include AI-related activities (see below, Part IV.4).

---

*Technologien im Spannungsfeld von Datenschutzrecht und BGB* (Mohr Siebeck 2020), 64 et seqq. and 128 et seqq.; and the discussion below.
[50] See, e.g., Bell and others, *Environmental Law* (9th edn, Oxford University Press 2017), 764 et seqq.
[51] Established by Directive 2003/87/EC of the European Parliament and of the Council of 13 October 2003 establishing a system for greenhouse gas emission allowance trading within the Union, OJ L 275, 25.10.2003, p. 32, as amended.
[52] Directive 2000/60/EC of the European Parliament and of the Council of 23 October 2000 establishing a framework for Community action in the field of water policy, OJ L 327, 22.12.2000, p. 1.
[53] Cf. Bell and others, op. cit. *supra* note 58, 27.
[54] Ibid, 764 et seqq.
[55] See, e.g., Hintermayer, "A carbon price floor in the reformed EU ETS: Design matters!" (2020) 147 *Energy Policy Article* 111905, 1; von Landenberg-Roberg, "Transformation durch innovationsfördernde Regulierung" (2023) *Zeitschrift für Umweltrecht* 148, 152; Fisher and others, *Environmental Law: Text, Cases & Materials* (Oxford University Press 2013), 617. The Industrial Emissions Directive 2010/75/EU, and its interplay with the ETS, will be bracketed here for reasons of scope and limited practical importance; see, e.g., Bell and others, op. cit. *supra* note 58, 495 et seqq. and 554.
[56] von Landenberg-Roberg, op. cit. *supra* note 65, 154.
[57] Fisher and others, op. cit. *supra* note 65, 617.

### b) The Water Framework Directive

A comparable outcome is observed regarding water usage. While there are no EU regulations explicitly circumscribing the water consumption attributed to AI and emerging technologies, broader water management and non-pollution policies, as stipulated by Article 4 of the EU Water Framework Directive (WFD), can come into play. Specifically, large data centres employing liquid cooling mechanisms may be subject to these overarching regulations, necessitating the adoption of sustainable water management practices.

During the AI production cycle, water–for the greatest part clean freshwater–is both withdrawn and consumed for cooling servers, electricity generation, and manufacturing.[58] Water withdrawal (also called "water use") designates the removal of water from ground or surface water sources for any uses, e.g., in agricultural, industrial or municipal settings.[59] Consumption is defined as the difference between withdrawal and discharge;[60] hence, it refers to any water that is removed from the immediate water environment, for reasons such as evaporation or incorporation into products and plants.[61] For example, in 2022, Google's own data centres withdrew 25 billion and consumed almost 20 billion need litres of water for on-site cooling, mostly (77%) potable water.[62] Typically, around 80% of withdrawn water is consumed by data centres due to evaporation.[63]

Consumed water is not available for any downstream uses.[64] But even water merely withdrawn reduces the amount of freshwater immediately available for other uses, such as irrigation or drinking water supply, due to the finite nature of water sources; and it often has to be retreated to make it fit for other uses, particularly drinking water supply.[65] Hence, both withdrawal and consumption contribute to water stress; but consumption is more important for scarcity.[66]

These effects are only indirectly addressed by the WFD. The Directive covers a variety of water bodies, including inland surface waters such as rivers and lakes, transitional waters like estuaries, coastal waters, and groundwater reserves. Articulated under Article 9 of the WFD, the "polluter pays" principle is employed as a strategy to mitigate pollution. Industries such as agriculture, manufacturing, water supply, and wastewater treatment are notably impacted by this Directive, given their potential roles either as contributors to water pollution or as sectors highly dependent on water resources. AI and emerging technologies primarily raise questions of water quantity rather than pollution, as discussed.

The WFD does not introduce caps for water usage, though, as the ETS does for GHG emissions; it primarily focuses on water quality instead.[67] To date, the data does not show that water consumed for AI training is, by any means, more polluting than any other industrial water usage; however, consumed water is not available for downstream uses, including drinking water supply. The Directive does, nevertheless, treat water quantity as an ancillary element to water quality (Recital 19 WFD). Hence, some parts of the WFD aim to specifically protect the quantity of available water, for example, through an economic analysis of water use (Art. 5(1) WFD); and monitoring the volume of surface and groundwater (Art. 8(1) WFD).[68] Furthermore, the quantity of water is an important element of the ecological status of surface water, like rivers and lakes, and groundwater.[69] Significantly, the CJEU has held that any deterioration of only one aspect of the water status violates the legally binding obligation to prevent the degradation of water status (Art. 4(1)(a) WFD), even if the overall status of the water element does not change.[70] This means that quantity degradation can be assessed and prosecuted separately[71] even if decreasing quantity is "offset" overall by enhanced water quality parameters.[72] As the water status regime applies to individual projects,[73] the water consumption by, for example, data centres can and must be taken into account in the authorisation process. Nonetheless, the framework lacks binding targets comparable to the ETS.

---

[58] Li and others, , op. cit. *supra* note 6, 3-4; the examples correspond to scope 1, scope 2 and scope 3 uses.
[59] Reig, "What's the difference between water use and water consumption?", *World Resources Institute Commentary* (12 March 2013).
[60] Macknick et al., A review of operational water consumption and withdrawal factors for electricity generating technologies, NREL Tech. Report: NREL/TP-6A20-50900, 2011, 2.
[61] Reig, op. cit. *supra* note 69.
[62] Google, Environmental Report, 2023, <sustainability.google/reports/>, 49 et seq., 95, (last visited 2 Dec 2023).
[63] Li and others, op. cit. *supra* note 6 .
[64] Ibid, 3.
[65] Reig, op. cit. *supra* note 69.
[66] Reig,op. cit. *supra* note 69What's the difference between water use and water consumption? World Resources Institute Commentary, 2013, https://www.wri.org/insights/whats-difference-between-water-use-and-water-consumption.
[67] See Recital 25 Directive 2000/60/EC of the European Parliament and of the Council establishing a framework for Community action in the field of water policy (Water Framework Directive) O.J. 2000, L 327/1; , Fisher, Lange and Scotford *Environmental Law: Text, Cases & Materials,* 471 (OUP, 2013); Bell and others, *Environmental Law,* 627, 8th ed. (OUP, 2013).

In summary, although EU environmental legislation has not been explicitly tailored to address AI and emerging technologies, it does exert indirect regulatory constraints through established frameworks like the ETS and the WFD. The latter, particularly, already require taking the water consumption of any facility, including data centres for AI, into account during the authorisation process, as discussed. The ETS, on the other hand, only very partially maps onto the challenges generated by AI due to its sectorial framework. Given the expanding environmental impact of AI, there is an increasing urgency for the EU to modify its current environmental statutes to more directly cover these emerging technologies (see Part IV.).

## 2. The GDPR

Another general regulatory framework applying to AI is the GDPR. In this context, environmental considerations have not been thoroughly explored. As I shall argue, the GDPR does not stand in the way of using AI in general, but an environmentally-aware interpretation may mitigate some of its negative environmental effects. However, counterintuitively, this may, in some cases, limit the reach of some subjective rights long taken for granted, such as the right to erasure.

At a general level, there are two ways in which data processing operations may incur environmental costs related to climate change: via direct and indirect environmental costs. First, processing operations can contribute to climate change by consuming energy and water, thereby increasing GHG emissions and contributing to resource scarcity (*direct environmental costs* of data processing). A prominent example is the energy-intensive training and deployment of AI systems. To the extent personal data are processed, the GDPR might take environmental costs into account and limit data processing accordingly. As we shall see, key concepts and rights of the GDPR can be re-construed to facilitate an environmentally aware interpretation of the GDPR.

---

[68] Bell and others, *Environmental Law,* 627, 8th ed. (OUP, 2013).
[69] See Annex V O.J. 2000, L 327/ 1 (Water Framework Directive), for example: point 1.1.1. (rivers); point 1.1.2. (lakes); point 2.1 (groundwater quantitative status).
[70] CJEU, Case C-461/13, Bund für Umwelt und Naturschutz Deutschland eV v Bundesrepublik Deutschland (Weser), EU:C:2015:433, para. 70.
[71] More specifically, water quantity is an important element of the hydromorphological quality of lakes and rivers (Annex V O.J. 2000, L 327/1 (Water Framework Directive), points 1.2.1. and 1.2.2.), see also Fisher, Lange and Scotford, *Environmental Law: Text, Cases & Materials,* 477 (OUP, 2013); there is a separate regime for groundwater quantity, see n 69.
[72] Cf. Bell and others Environmental Law, 629.
[73] CJEU, Case C-461/13, Bund für Umwelt und Naturschutz Deutschland eV v Bundesrepublik Deutschland (Weser), EU:C:2015:433, para. 47-51.

Second, even data processing that is not energy intensive may trigger environmental costs down the road. These could be labelled *indirect environmental costs* of data processing. Such costs may arise via the GHG emitted through the lifecycle of the machinery needed to conduct the processing, such as laptops, IoT devices or industrial machinery. Furthermore, data analysis may be conducted for purposes that generate environmental costs. For example, any processing of personal data by an organisation actively denying climate change and fighting against climate-conscious policy measures indirectly contributes to global warming. How should personal data processing for an email list of climate deniers, or for optimizing the construction of coal plants, be considered under the GDPR?

The incorporation of environmental sustainability considerations into the interpretation of the GDPR presents an intricate, yet uncharted, challenge for legal scholars and policymakers. As it stands, the GDPR does not enumerate any specific rules or guidelines that directly link data protection with environmental sustainability. This is not surprising given the strong focus of the GDPR on the relationship between individual data subjects and controllers. Even evidently relevant third parties are left outside the scope of the GDPR–the obligations of data protection by design and default under Article 25 GDPR, for example, do not even apply to manufacturers of data processing devices (see Recital 78 GDPR). Neither do sustainability considerations feature explicitly under any of the established principles for data processing (Art. 5 GDPR), nor is there a concrete limit on data processing operations based on environmental criteria. However, there are several provisions in the GDPR which, by virtue of their open wording or their reference to third-party interests, may be interpreted as guardrails addressing questions of sustainability and environmental costs.

### a) Legitimate interests and purposes

As a first interpretive inroad into the narrow focus of the GDPR on the dyadic relationship between data subject and controller, one may turn to questions of *legitimate* processing. Such wording may be harnessed to build a link between specific types of processing and legal or societal values outside of the GDPR proper. For example, the principle of purpose limitation highlights that personal data may only be collected and processed for specific, explicit and legitimate purposes (Art. 5(1b) GDPR). Similarly, the balancing test under Article 6(1)(f) GDPR allows only legitimate interests of the controller to be considered when assessing whether processing has a legal basis in the absence of consent (or other legal grounds listed in Art. 6 GDPR).

To start with, it is generally agreed upon that illegal purposes and interests are, by extension, illegitimate.[74] Hence, if an activity, such as the construction of an illegal coal mine, runs afoul of environmental regulation, any personal data processing relating to this activity would be considered illegitimate in its purpose and interest. However, most environmentally damaging activities and purposes, from email lists fostering climate denialism to the construction of a database of workers helping to build a heavily GHG-emitting but approved coal power plant, are generally not illegal, at least not per se.

Thus, the question arises to what extent environmental costs can be factored into the analysis of legitimate purposes and interests. As the Working Party acknowledges, the notion of legitimacy is dynamic: it "can also change over time, depending on scientific and technological developments, and changes in society and cultural attitudes."[75]

---

[74] AG Bobek, Opinion of Dec. 19, 2018, Case C-40/17 *Fashion ID GmbH & Co.KG v. Verbraucherzentrale NRW e.V.*, EU:C:2018:1039, para. 122; Article 29 Data Protection Working Party, Opinion 06/2014 on the notion of legitimate interests of the data controller under Article 7 of Directive 95/46/EC, WP 217, 2014, 25.

As our data-driven society continues to expand, the environmental footprint of digital infrastructure is growing concomitantly–as is environmental awareness among the general public and the recognition that global warming and climate change must be mitigated. Taking up the dynamic concept of legitimacy championed by the Working Party, and recognizing the environment as a stakeholder, the GDPR would align itself with the broader EU agenda of sustainable development. Such an interpretation would, arguably, set a doctrinally justifiable precedent broadening the traditional ambit of data protection law to address the exigent challenges posed, e.g., by climate change.

Since any electronic data processing has direct and potentially indirect environmental costs, this mere fact alone cannot be enough to render purchasing or interests illegitimate. However, legitimate purposes and interests cannot be identical with their legality, either, as the framers could have plainly stated that fact, but chose the wider term "legitimate" instead. As a compromise avoiding both extremes, I suggest taking environmental costs into account in the following way in this context.

### i.  Direct environmental costs

Direct environmental costs, such as the energy and water consumption of processing, should be a factor in determining legitimacy. Arguably, a cost-benefit analysis could be employed, whereby only those activities with clearly excessive environmental costs relative to their societal benefit would be deemed illegitimate. This allows for the differentiation between typical computer-based operations, which are generally considered legitimate due to their ubiquitous role in modern society, and specialised activities such as energy-intensive AI training. The latter would still be seen as legitimate if they serve a socially beneficial purpose, like advancing scientific research or optimizing renewable energy production, thus balancing environmental impact against societal gains.

Borderline cases present a more intricate scenario for evaluation. For instance, highly energy- and water-consuming operations related to entertainment, as seen in AI and Metaverse applications, might raise concerns. These activities are generally considered 'legitimate purposes' under data protection norms, and they are not inherently malign. They represent an area where more detailed regulatory guidance seems necessary, potentially requiring a more nuanced definition of 'social benefit' that includes considerations of cultural or recreational value against their environmental cost. On the other end of the spectrum, activities like AI training specifically designed for unethical or harmful purposes, such as cyberattacks, would be deemed illegitimate, regardless of whether they meet a technical criminality threshold.[76] Such an approach would transcend legalistic considerations to incorporate ethical and environmental criteria in the assessment of data processing legitimacy.

### ii.  Indirect environmental costs

Concerning indirect environmental costs (e.g., supply chain; purpose of processing), the expected societal harm resulting from the activity should be considered, particularly with respect to the likelihood and severity of the harm. It bears noting that the environment does not factor as one of the risks described under Recital 75 GDPR as a typical risk of data processing. However, that list is not exhaustive and novel societal concerns may, arguably, be designated risks of data processing. Importantly, even perfectly addressing the risks named in Recital 75 GDPR becomes irrelevant once climate change makes human flourishing and the enjoyment of fundamental rights free from those risks impossible.

---

[75] Opinion 06/2014 on the notion of legitimate interests of the data controller under Article 7 of Directive 95/46/EC, WP 217, 2014, 25 fn. 48, available at <ec.europa.eu/justice/article-29/documentation/opinion-recommendation/files/2014/wp217_en.pdf>.

[76] If the activity is indeed criminalized, and the act conducted or at least attempted in a legally relevant way, AI training for that purpose may also count as "aiding and abetting" that crime, rendering the data processing straightforward illegitimate.

However, the GDPR can hardly "police" any secondary environmental effects of data processing. This would overburden agencies and overstretch the regulation's reach. Therefore, indirect environmental costs should only factor into, e.g., the balancing test if there is a clear, relevant, and not merely hypothetical link between the processing and the secondary environmental costs.

Such a link does not exist in the case of the email list of climate deniers mentioned above. However, it may be shown if an AI system is trained on personal data to optimise construction procedures for heavily GHG-emitting coal plants. Such infrastructure produces carbon "lock-in effects" with significant impact on climate change.[77] While an approval of the plant, via a comprehensive regulatory process, would potentially shield this activity from the critique of illegitimacy under the duty of sincere cooperation enshrined in Art. 4(3) TEU and spelled out by the CJEU in *Meta v. Bundeskartellamt*,[78] this is arguably different if the procedure did not even contemplate environmental costs. In these cases, indirect environmental costs of data processing clearly and relevantly linked to the data analysis can be one factor speaking in favour of the illegitimacy of the purposes or interests, even though they will rarely be sufficient to render the entire processing illegitimate on their own.

### b) Third-party interests in the balancing test

As indicated, the balancing test of Article 6(1)(f) GDPR presents itself as another potentially crucial point of entry for the collective dimension of sustainability considerations. The test itself is a key legal basis in machine learning contexts as consent will often be difficult to obtain from the large number of data subjects affected by big data and AI analyses. According to the case law of the CJEU, the three cumulatively necessary criteria must be met for data processing to be legal under the balancing test:

"first, the pursuit of a legitimate interest by the data controller or by the third party or parties to whom the data are disclosed; second, the need to process personal data for the purposes of the legitimate interests pursued; and third, that the fundamental rights and freedoms of the person concerned by the data protection do not take precedence."[79]

Under these conditions, the consideration of justified third-party interests is a complex issue.[80] As it stands, the existing legal language only allows these interests to favour data processing under the first criterion, rather than count against it under the third criterion. In its opinion on the notion of legitimate interest, the Article 29 Working Party succinctly stated that "is the broader stake that a controller may have in the processing, or the benefit that the controller derives - *or that society might derive* - from the processing".[81]

---

[77] <www.ipcc.ch/report/ar6/wg3/downloads/report/IPCC_AR6_WGIII_SummaryForPolicymakers.pdf>, C.4, (last visited 21. Dec. 2023).
[78] Cf. CJEU Case C-252/21, *Meta v. Bundeskartellamt*, EU:C:2023:527, para. 52-56 (concerning the departure of a competition authority from an assessment made by a data protection authority regarding GDPR violations).
[79] CJEU Case C-13/16, *Rīgas satiksme*, EU:C:2017:336, para. 28 (on the equivalent provision of Article 7(f) of the Data Protection Directive); confirmed in CJEU Case C-40/17, *Fashion ID GmbH & Co.KG v. Verbraucherzentrale NRW e.V.*, EU:C:2018:1039, para. 95.
[80] See also, in greater detail, Hacker, *Datenprivatrecht: Neue Technologien im Spannungsfeld von Datenschutzrecht und BGB* (Mohr Siebeck 2020), 273 et seqq.; see also Barocas and Levy, "Privacy dependencies" 95 Wash L Rev 555 (2020)

This creates an undue imbalance, however: users with lax data protection preferences who are willing to permit tracking for convenience or budgetary considerations would have their interests accounted for, while negative externalities for third parties, such as privacy-intrusive inferences[82] or climate effects, could be overlooked. Such an approach appears inconsistent when interpreted in alignment with primary law and in the context of Article 20 of the Charter of Fundamental Rights (CHF).[83] Specifically, no compelling reason can be found for taking third-party interests only into account for overriding the fundamental rights and freedoms of the data subject, but not to protect the core interests and rights of affected parties. Hence, in the light of Article 20 ChFR, the balancing test must be interpreted such as to allow for the consideration of third-party interests within the third criterion as well: If a shift towards incorporating third-party interests is to occur, it must be applied in a balanced, or symmetric, fashion.

This perspective is increasingly supported by the CJEU's case law concerning the direct horizontal effect of the Charter's fundamental rights.[84] While the CJEU has only ruled on the direct horizontal effect of specific Charter rights such as Art. 21 (non-discrimination) and Art. 31(2) (fair and just working conditions), the jurisprudence can likely be extended to data protection under Article 8 ChFR. The precise extent to which indirect impacts, such as adverse inferences,[85] i.e., deducing traits detrimental to the data subject based on data revealed by other, similarly situated parties, could actually infringe upon the data protection rights of third parties remains an area for rigorous academic inquiry. For example, empirical and theoretical studies suggest that privacy may unravel if individuals have strong incentives to disclose positive attributes, leading to a presumption of negative traits for those unwilling to disclose their attributes.[86] However, as a minimum consequence of the referenced CJEU jurisprudence, the impact on third-party data protection rights should be taken into account within private law contexts, particularly in the scope of Article 6(1)(f) GDPR. This provision thus emerges as a crucial entry point for considering the collective dimensions of data protection.

---

[81] Article 29 Data Protection Working Party, Opinion 06/2014 on the notion of legitimate interests of the data controller under Article 7 of Directive 95/46/EC, WP 217, 2014, 24, available at <ec.europa.eu/justice/article-29/documentation/opinion-recommendation/files/2014/wp217_en.pdf>.

[82] See, e.g., Barocas and Levy, op. cit. *supra* note 93.

[83] See Schweitzer, 'Neue Machtlagen in der digitalen Welt? Das Beispiel unentgeltlicher Leistungen' in Körber and Kühling (ed), *Regulierung-Wettbewerb-Innovation* (Nomos 2017), pp. 269 - 306at 282; Hacker, op. cit. *supra* note 93, at 274.

[84] See, e.g., CJEU Case C-414/16, *Egenberger v. Evangelisches Werk für Diakonie und Entwicklung e.V.*, EU:C:2018:257, para. 76; Case C-68/17, *IR v. JQ*, EU:C:2018:696, para. 69; Joined Cases C-569/16 & 570/16, *Bauer und Willmeroth*, EU:C:2018:871, para. 89-90 and 92; Case C-193/17, *Cresco Investigation v. Achatzi*, EU:C:2019:43, para. 79 et seq.; see also, for the long-standing debate, Lenaerts and Gutiérrez-Fons, "The constitutional allocation of powers and general principles of EU law" 47 CML Rev. (2010) 1629-1669, particularly 1648; Frantziou, "(Most of) the Charter of Fundamental Rights is Horizontally Applicable: ECJ 6 November 2018, Joined Cases C-569/16 and C-570/16, Bauer et al" 15 EUConst (2019), 306-323; Frantziou, "The Horizontal Effect of the Charter of Fundamental Rights of the EU: Rediscovering the Reasons for Horizontality" 21 ELJ (2015), 657-679; Ciacchi, "Egenberger and Comparative Law: A Victory of the Direct Horizontal Effect of Fundamental Rights" 5 *European Journal of Comparative Law and Governance* (2018), 207-211; for a more critical take, see Fornasier, "The impact of EU fundamental rights on private relationships: direct or indirect effect?" 23 E.R.P.L. (2015), 29-46, at 32 et seqq.

[85] See Barocas and Levy, op. cit. *supra* note 93, at 599 et seqq.

[86] See e.g., Peppet, "Unraveling privacy: The personal prospectus and the threat of a full-disclosure future" 105 Nw U L Rev (2011), 1153-1203; Hermstruwer, "Contracting around privacy: the (behavioral) law and economics of consent and big data" (2017) 8 J Intell Prop Info Tech & Elec Com L 9, para. 21 et seqq.

In our context, however, the important question is: do the interests of the parties that must be considered under the third criterion also include a collective interest in climate change mitigation? On the one hand, as in the case of legitimate purposes, one could argue that the GDPR is not the right locus to integrate societal concerns surrounding a warming planet. Under this reading, only specific data protection risks should qualify in the balancing test. In my view, however, three reasons speak in favour of considering sustainability interests. First and foremost, the "equal arms" argument applies again: if legitimate interests of the controller expand clearly beyond mere data processing (see, e.g., Recital 47 GDPR: marketing), this extension must be mirrored on the data subject's side. Second, sustainability interests are not exogenous, but inherent in data processing, as the computer science overview has shown. As mentioned, the fact that they are not named in Recital 75 cannot be dispositive given its non-exhaustive character. Finally, while climate change affects third parties in differential ways, a warming planet is ultimately detrimental to the health and well-being of most EU citizens given its overall negative direct (e.g., extreme weather events[87]) and indirect consequences (e.g., geopolitical and social tensions;[88] economic costs[89]). Environmental costs should therefore count in the balancing test.

As in the case of legitimate purpose or interest, this does not mean that any direct or indirect environmental costs of data processing immediately disqualify it under the balancing test. However, the climate effects of processing–based on the expected GHG emissions, water consumption, and other pertaining and readily calculable metrics–should be one factor to be considered in the balancing exercise. Particularly climate-unfriendly forms of processing, such as the training of large neural networks, will face a greater burden of justification. For example, one could again argue that training or deploying AI merely for entertainment purposes should face particularly strict scrutiny, and could–in extreme cases–even be disqualified under the balancing test primarily because of staggering environmental costs.

### 3. Subjective rights and environmental costs

As the previous discussion has shown, sustainability emerges as a collective interest of legal relevance in data protection law. Significantly, it may also reconfigure certain existing rights we have come to take for granted. While the consideration of environmental costs, at other points, may restrict AI development and deployment, the following perspective on individual rights imply that environmental considerations may also limit data subject rights. On a theoretical level, this question links to the broader issue of the value and position of collective interests in law.[90] From an economic perspective, sustainability and climate change mitigation qualify as a public good.[91] This not only points to the importance of regulatory intervention for its implementation but also raises important doctrinal questions concerning the relationship to subjective rights: using several examples drawn from data protection and non-discrimination law, the following sections explore to what extent existing subjective rights may have to be limited, or buttressed, by sustainability considerations.[92] Taken together, these examples again point to the larger question of the relevance and status of collective interests in sustainability in a body of the law that, in private law particularly, has for centuries been understood as structuring largely bilateral relationships.

---

[87] Stott, "How climate change affects extreme weather events" *Science* , 352 (2016), 1517.
[88] Dalby, "The geopolitics of climate change" *Political Geography,* 37 (2013), 38; Nordhaus, "The climate club: how to fix a failing global effort", *Foreign Aff.* 99 (2020).
[89] Stern, "The economics of climate change", *American Economic Review* 98 *(*2008) *,* 1; Rennert and others, "Comprehensive evidence implies a higher social cost of CO2" *Nature,* 610 (2022), 687.
[90] See, e.g., Newman, "Collective interests and collective rights", *The American Journal of Jurisprudence,* (2004) *49, 127*; Mulcahy, "The collective interest in private dispute resolution", *Oxford Journal of Legal Studies,* (2013) *33,* 59; See, e.g., Hacker, *Verhaltensökonomik und Normativität: die Grenzen des Informationsmodells im Privatrecht und seine Alternativen* (Mohr Siebeck, 2017), 368 et seqq.
[91] See, e.g., Stern, "The economics of climate change" (cambridge University press), 2007.; Wagner and Weitzman, *Climate shock* (Princeton University Press, 2016).

### a) Erasure versus sustainability

To start with, we shall consider the "right to be forgotten" (Article 17 GDPR), which allows data subjects, under certain conditions, to request the erasure of their personal data. Now imagine that a large AI model was trained on supposedly anonymised medical data and is used for cancer detection. Some individuals whose data were contained in the training data set may be re-identified with novel technical tools (cf. Recital 26).[93] One of them exercises her right to erasure under Article 17(1)(a) GDPR. As a consequence, not only may her data points have to be deleted from the training data, but the entire AI model may have to be re-trained[94]–entailing significant GHG emissions.

It is submitted that the subjective right to erasure, in such situations, has to be balanced against the collective interest in mitigating climate change. This ties in with a more holistic view of the different rights and protection regimes in EU law. In environmental law, for example, the CJEU has ruled that overriding public interest, for example in the provision of renewable energy, may justify the deterioration of the quality of waters.[95] Doctrinally, in the context of the GDPR, such a "sustainability limitation" might be based on Art. 17(3)(c) or (d) GDPR. According to the former, the right to erasure does not exist if processing is necessary for reasons of public interest in the area of public health, provided that the constraints of Article 9(2)(h) or (i) and Article 9(3) GDPR are heeded. This requires, importantly, a specific law under EU or Member State law authorizing the processing of sensitive data in the specific case and providing adequate safeguards.

While such a law would indeed be a favourable course of action, the question remains whether, in its absence, a possible defence against the erasure request in this and other cases might be based on Article 17(3)(d) GDPR. Pursuant to this provision, the erasure request may be denied if the processing is necessary for "for archiving purposes in the public interest, scientific or historical research purposes or statistical purposes in accordance with Article 89(1) [GDPR] in so far as [the erasure] is likely to render impossible or seriously impair the achievement of the objectives of that processing". Art. 17(3) GDPR establishes an exemption whose validity, in each single case, can only be ascertained by striking an equitable balance between the mentioned rights and interests.[96]

Based on a grammatical interpretation of Article 17(3)(d) GDPR, one will hardly find that the erasure request will seriously impair reaching the objectives of any research purpose inherent in building the model: the deletion of a single data point from the training data set will rarely have any noticeable effect. One could potentially argue that taken together, a multitude of erasure requests could have such an effect; but the extent to which such extrapolations might be considered in evaluating a single erasure request–particularly if the other requests have not yet been formulated–is unclear. Hence, the woman from the example stands a decent chance of seeing her erasure request granted under a literal interpretation of Article 17(3)(d) GDPR.

---

[92] The balancing exercise actually involves the following rights and interests: those of the data subject; of the controller; the interest in AI development and the concrete deployment; and sustainability. Due to space constraints, the following considerations focus on the first and the last aspect.
[93] See, e.g., Finck and Pallas, "They who must not be identified—distinguishing personal from non-personal data under the GDPR," *International Data Privacy Law* 10.1 (2020); Hacker, "A legal framework for AI training data—from first principles to the Artificial Intelligence Act", *Law, Innovation and Technology* 13.2 (2021).
[94] See, e.g., Michael Veale, Reuben Binns and Lilian Edwards, 'Algorithms that remember: model inversion attacks and data protection law' (2018) *376 Philosophical Transactions of the Royal Society A: Mathematical, Physical and Engineering Sciences* 20180083; Tiffany C Li, 'Algorithmic Destruction' (2022) *75 SMU L Rev* 479; Alessandro Achille and others, 'AI Model Disgorgement: Methods and Choices' (2023) *arXiv preprint arXiv*:230403545.
[95] CJEU, Case C-346-14, Commission v Austria (Schwarze Sulm), para. 71 and 83, based on Article 4 (7) WFD.
[96] See, e.g., CJEU, Case C-398/15, Manni, para. 57-63.

Arguably, however, Article 17(3) GDPR only imperfectly sketches and operationalises a constitutionally required balancing exercise. As the CJEU spelled out in the infamous *Google Spain* case, the "right to be forgotten" requires a comprehensive assessment of the rights and interests of the respective parties to determine whether any erasure request must be granted.[97] Importantly, the Court in *Google Spain* specifically included the societal impact of technologies into the balancing.[98] The CJEU has taken up this balancing exercise more recently in the *Manni* case, for example, albeit in the context of a legal obligation to process data (company register).[99] Nonetheless, I submit that a purposive, teleological interpretation of Article 17(3) GDPR must not be limited to the cases enumerated in letters (a) to (e). Rather, an unwritten exemption should be read into Article 17(3) GDPR according to which erasure requests may be denied if a comprehensive balancing exercise including the constitutionally protected rights and interests of the involved parties finds a preponderance of the rights and interests of the data controller.

Importantly, Article 17(3)(c) and (d) GDPR demonstrate that collective interests, such as public health, archiving, or research, may limit individual erasure requests. This links the right to erasure to a growing debate on the relevance of third-party interests for interpreting the GDPR.[100] That debate, however, has so far focused on the extent to which disclosure of individual data points, and their processing, may generate (negative) externalities to third parties; for example, ML models might inference their hitherto undisclosed preferences by aggregating enough data from similarly situated persons (triangulation).[101] The impetus, then, would be to restrict data processing in the interest of the data subjects whose preferences might be the object of inference.

The question of sustainability puts this issue on its head: under this perspective, we must ask if the (climate) externalities of the exercise of individual data subject rights may effectively limit the reach or even existence of these individual rights. Again, as Article 17(3)(c) and (d) GDPR suggest, such reasoning is not foreign to the GDPR and its right to erasure. Framed as a constitutionally required weighting exercise, it involves balancing the right of data protection against the right to, or at least interest in, saving energy and reducing GHG emissions for the sake of climate change mitigation. Importantly, while the latter is a collective interest, precisely the debate around data externalities shows that the right of data protection need not be understood as a merely individual right, but that it also embodies a collective dimension.

Doctrinally speaking, however, the Charter does not contain individual rights to a healthy environment.[102] Article 37 ChFR lists environmental protection only as a principle, not an individual, justiciable right.[103] However, the analysis hardly ends here. Climate change is already, by all scientific standards, affecting the rights to property (Article 17(1) ChFR), health and physical integrity (Article 3 ChFR), and life (Article 2 ChFR) for a growing number of persons protected by the Charter. These rights support the freedom to conduct the business under Article 16 ChFR and have to be balanced against the erasure applicant's right to data protection flowing from Article 8 ChFR. Such a reading is in line with the number of European cases in which litigants have, by invoking such individual rights,[104] successfully challenged climate-related laws and practices in front of the Dutch courts[105] and the German Federal Constitutional Court.[106]

---

[97] CJEU, Case C-131/12, Google Spain, para. 76 and 81.
[98] CJEU, Case C-131/12, Google Spain, para. 80.
[99] CJEU, Case C-398/15, Manni, para. 63.
[100] See, e.g., Barocas and Levy, "Privacy dependencies"; Ben-Shahar, "Data Pollution" *Journal of Legal Analysis* 11 (2019), 104; Hacker, *Datenprivatrecht* (Mohr Siebeck, 2020), 64 et seqq.
[101] Zarsky, "Desperately seeking solutions: using implementation-based solutions for the troubles of information privacy in the age of data mining and the internet society" 56 Me L Rev (2004), 13, 43 et seq.; Fuster and others, "Predictably unequal? The effects of machine learning on credit markets" *The Journal of Finance,* 77 (2022) 5, 15 et seqq.
[102] Josephine Van Zeben, 'The Role of the EU Charter of Fundamental Rights in Climate Litigation' (2021) 22 *German Law Journal* 1499, 1505.

Ultimately, such cases will have to be resolved on a case-by-case basis. However, the guideline, in my view, should be that a clear discrepancy between the individual and collective harms flowing from the continued storage/use of the data point in question and the harms resulting from the climate costs of, e.g., retraining the model, should indeed lead to an exemption from the right of erasure under Article 17(3) GDPR. This will depend, inter alia, on the nature of the personal data (e.g., its proximity to sensitive data and its importance for the exercise of other fundamental rights); the extent of the projected environmental costs; potential alternatives the data controller might have used to minimise climate costs in the case of erasure requests (e.g., machine unlearning[107] or sharding[108]). The social utility of the model itself, however, should not factor into the equation as the erasure will generally not lead to the deletion of the model, but only its (partial) retraining;[109] if, indeed, the purpose of the model is defeated by the erasure request, Article 17(3)(d) GDPR provides the more specific norm for the balancing exercise.

In our hypothetical, if indeed the GHG emissions from retraining model are significant (e.g., the yearly emissions of a medium-sized town) there is arguably a case to be made that the erasure request should, despite the sensitive nature of the data, eventually be denied.

### b) Transparency versus sustainability

Besides erasure, transparency and access requests constitute another domain of subjective rights potentially challenged by sustainability considerations. Transparent models may, in some cases, be more resource intensive than opaque "black-box" models. This holds particularly if a post-hoc explanation algorithm must be deployed on top of the machine learning model. Such an algorithm analyses the fully trained AI model to provide certain types of reasons why a certain output was achieved or how the model functions overall.[110] Particularly powerful models using deep learning, such as neural networks, often can only be explained post hoc.[111] Conversely, the use of simpler models that are interpretable from the start may indeed reduce energy consumption.

---

[103] Elisa Morgera and Gracia Marín Durán, 'Article 37–Environmental Protection' in Tamara Katherine Hervey, et al. (ed), *The EU Charter of Fundamental Rights* (Hart/Nomos 2022), 1041.
[104] See, in greater detail, Van Zeben, 'The Role of the EU Charter of Fundamental Rights in Climate Litigation', 1501 et seqq.
[105] Hof's-Gravenhage, October 9, 2018, Case-19/00135, Stichting Urgenda (Staat der Nederlanden/Stichting Urgenda); Rb. den Haag, May 26, 2021, Prg. 2021 mnt HA ZA 19-379 (Milieudefensie/Royal Dutch Shell PLC), para 5.3, https://uitspraken.rechtspraak.nl/inziendocument?id=ECLI:NL:RBDHA:2021:5339.
[106] Bundesverfassungsgericht, Case 1 BvR 2656/18, March 24, 2021, http://www.bverfg.de/e/rs20210324_1bvr265618en.html.
[107] Nguyen and others, "A survey of machine unlearning" (2022) arXiv preprint arXiv:220902299; Sekhari and others, "Remember what you want to forget: Algorithms for machine unlearning," *Advances in Neural Information Processing Systems,* 34 (2021), 18075; Cao and Yang, "Towards making systems forget with machine unlearning" 2015 IEEE Symposium on Security and Privacy, IEEE (2015), 463; Floridi, "Machine Unlearning: its nature, scope, and importance for a "delete culture" *Philosophy & Technology,* 36 (2023), *42*.
[108] Lucas Bourtoule and others, 'Machine unlearning' (2021) *2021 IEEE Symposium on Security and Privacy* (SP) 141.
[109] But see the references cited in n. 94.
[110] Alejandro Barredo Arrieta and others, 'Explainable Artificial Intelligence (XAI): Concepts, taxonomies, opportunities and challenges toward responsible AI' (2020) *58 Information Fusion* 82.

For instance, consider a scenario where only a black-box model can solve a difficult problem, such as detecting fraudulent transactions or diagnosing diseases. Suppose further that the law requires the model to provide explanations for its predictions, as recently held by the Amsterdam Court of Appeals in a case involving Ola and Uber;[112] and that a post-hoc explanation algorithm exists for this purpose. However, the post-hoc explanation algorithm is very resource intensive, as it needs to analyse the black-box model and generate explanations for each case. This may result in increased energy consumption and environmental impact. In such a situation, one might wonder whether the harm of opacity is less than the harm to sustainability.

That is, would it be better to use a simpler and more transparent model that is interpretable from the start, even if it sacrifices some accuracy? Or would it be preferable to use a more accurate but opaque model that requires a costly post-hoc explanation method? How can we balance these conflicting objectives and values? These questions are not easy to answer, as they depend on various factors, such as the nature and importance of the problem, the availability and quality of the data, the expectations and preferences of the users and stakeholders, and the legal and ethical implications of the decisions.

Doctrinally speaking, a balancing exercise will have to be undertaken under the access provision (e.g., Article 15(1)(h) GDPR) that is similar in nature and scope to the one discussed under Article 17 GDPR. As a result, transparency may be limited to an aggregate statement about the whole model, instead of individual cases, if this is computationally less resource-intensive; to an approximation; or may in extreme cases even be entirely dispensed with, depending on the outcome of the balancing exercise.

### c) Non-discrimination versus sustainability

In ways similar to post hoc explanations, post-hoc fairness algorithms may mitigate bias in machine learning systems.[113] Just like post-hoc explanations are added once the model has delivered an output, these fairness interventions correct the results of an algorithmic outcome to mitigate bias.[114] However, these methods may have unintended consequences for the environment and sustainability, too. Adding post-hoc fairness algorithms to the original algorithmic process may again increase the energy consumption and carbon footprint of the system, as they require additional computational resources and data processing.

---

[111] Ibid, 95; Zachary C Lipton, 'The mythos of model interpretability: In machine learning, the concept of interpretability is both important and slippery' (2018) 16 *Queue* 31.

[112] See Amsterdam Court of Appeals, Case 200.295.806/01, April 4, 2023, https://uitspraken.rechtspraak.nl/#!/details?id=ECLI:NL:GHAMS:2023:804; an unofficial translation can be found here: <https://5b88ae42-7f11-4060-85ff-4724bbfed648.usrfiles.com/ugd/5b88ae_de414334d89844bea61deaaebedfbbfe.pdf>; see also Jakob Turner, "Amsterdam Court Upholds Appeal in Algorithmic Decision-Making Test Case: Drivers v Uber and Ola" *Fountain Court Blog* (June 4, 2023), <https://www.fountaincourt.co.uk/2023/04/amsterdam-court-upholds-appeal-in-algorithmic-decision-making-test-case-drivers-v-uber-and-ola/> (last visited December 21, 2023); cf. also Selbst and Powles, "Meaningful information and the right to explanation," *International Data Privacy Law,* 7 (2017), 233; Wachter, ́Mittelstadt and Floridi, "Why a right to explanation of automated decision-making does not exist in the general data protection regulation" *International Data Privacy Law*, 7 (2017), 76; Malgieri and Comandé, "Why a right to legibility of automated decision-making exists in the general data protection regulation," *International Data Privacy Law,* 7 (2017), 243.

[113] See, e.g., Pessach and Shmueli, "A review on fairness in machine learning" 55(3) CSUR (2022), Article 51, 10 et seqq.; Zehlike, Hacker and Wiedemann, "Matching code and law: achieving algorithmic fairness with optimal transport," *Data Mining and Knowledge Discovery* 34 (2020), 163.

[114] See, e.g., Petersen and others, "Post-processing for individual fairness" *Advances in Neural Information Processing Systems 34* (2021*),* 25944; Pessach and Shmueli, "A review on fairness in machine learning," 55(3) CSUR (2022)
.

However, between opacity and bias, I would argue that non-discrimination is more important and urgent to ensure in algorithmic systems, as long as the energy consumption of the fairness system is reasonable and within acceptable limits. While it may be intellectually vexing to be faced with a non-transparent decision, discrimination entails a typically more serious, more concrete, economic as well as dignitary harm.[115] From a normative perspective, therefore, discrimination often has more severe and direct impacts on the affected persons than opacity. Lack of algorithmic fairness may lead to important, discriminatory life outcomes, such as denial of opportunities, resources, or benefits, or exposure to risks, harms, or disadvantages. Moreover, discrimination usually disproportionately affects those who are already marginalised or underserved by society, such as minorities, women, or people with disabilities.[116]

Moreover, fairness algorithms usually do not incur excessive GHG emissions.[117] Hence, the additional environmental cost will often be justified in view of the greater harm that discrimination inflicted upon individuals and society. This analysis is complicated, however, by the opportunity offered by increased transparency to detect discrimination in the first place.[118] Where this can indeed be expected, environmentally costly post-hoc disclosures will be generally justified even if they entail significant–but not unreasonable–environmental costs. Overall, non-discrimination is a fundamental right that should be respected and protected in algorithmic systems–even if climate costs are substantial.

### 4. The EU AI Act

In the general legal literature, a growing discussion exists about the interrelated impacts of climate change on the law, and vice versa.[119] However, for AI and technology regulation more specifically, questions of climate change and sustainability are still underexplored.[120] "The uptake of AI applications is beneficial for the environment," the Commission laconically concludes.[121] While AI indeed has a role to play here,[122] this analysis is dangerously one-sided and ignores recent developments in computer science reviewed above, suggesting a highly significant and rapidly growing contribution of AI, and ICT, to GHG emissions.

#### a) Voluntary commitments

This section of the article will therefore explore how existing or forthcoming legal instruments may render AI more sustainable. As a key example, let us turn to the EU AI Act. Under the General Approach adopted by the Council of the EU on December 6, 2022 (AI Act Council

---

[115] Onwuachi-Willig, "Reconceptualizing the harms of discrimination" (2019) *105 Virginia Law Review* 343, 349 et seqq.; Berndt Rasmussen, "Harm and discrimination," *Ethical Theory and Moral Practice 22* (2019), 873, 877 et seqq.

[116] See, e.g., Ian Ayres, "Fair driving: Gender and race discrimination in retail car negotiations" (1991) *Harvard Law Review,* 817; Buolamwini and Gebru, 'Gender shades: Intersectional accuracy disparities in commercial gender classification' (2018) *Conference on Fairness, Accountability and Transparency,* 77.

[117] See, e.g., the runtime detailed in Henzinger and others, 'Monitoring Algorithmic Fairness' (2023) arXiv preprint arXiv:230515979.

[118] See only Kleinberg and others, 'Discrimination in the Age of Algorithms' (2018) *10 Journal of Legal Analysis* 113, 152.

[119] See n. 19.

[120] But see references in n. 20.

[121] Proposal for a Directive of the European Parliament and of the Council and of the Council on adapting non-contractual civil liability rules to artificial intelligence (AI Liability Directive) COM/2022/496, p. 4.

[122] Cowls and others, op. cit. *supra* note 23.

Version),¹²³ the European legislator only encourages voluntary codes of conduct concerning environmental sustainability (Art. 69(2) AI Act Council Version). Such voluntary commitments are, however, hardly enough to mitigate climate change contributions in a key and burgeoning sector of technological and economic development, as the history of failed voluntary commitments to reign in climate change quite clearly shows.¹²⁴

This limited commitment changed on June 14, 2023, when the European Parliament adopted its position on the AI Act (AI Act EP Version¹²⁵). However, while the EP Version goes further in addressing environmental concerns, it still falls short of taking sufficient action. The amendments contain different sets of rules concerning sustainability. I will structure their analysis around five pillars: goals and principles; preferential access to research funding and sandboxes; information approaches; risk assessment; and guidance and review.

### b) Amendments by the European Parliament

The first EP amendment focuses on goals and principles. Article 1 AI Act EP Version sets out the general objectives of the regulation, including the prevention of harm to the environment. Furthermore, environmental sustainability figures prominently among the new principles for AI development and deployment (Article 4a(1)(f) AI Act EP Version). However, they arguably lack the necessary regulatory teeth to incentivise meaningful action. The Act, in these sections, sets lofty goals without providing concrete measures to ensure their implementation. If, eventually, the principles become part of the enforceable rules–such as the data protection principles in Article 5 GDPR–then indeed they may foster principle-based AI development. As such, they could qualify as backup rules offering a last resort in case an AI system circumvents specific AI Act rules, meeting requirements by the letter but violating them in spirit. Under the GDPR, recent administrative rulings against Meta show the potential of weaponizing mere principles in infringement proceedings.¹²⁶

The second pillar the EP suggests concerns funding and support. It proposes preferential access to research funding and sandboxes for AI systems promising to make a positive impact on the environment, as outlined in Article 54a. This provision aims to encourage the development of AI systems that prioritise environmental sustainability. However, further details and mechanisms for implementation are required to effectively promote these objectives; moreover, to develop truly ground-breaking AI systems in the range of foundation models, significantly more funding concerning the provision of compute infrastructure will be necessary.¹²⁷

Information approaches are emphasised in the third pillar. Article 12(2a) AI Act EP Version requires the measurement and calculation of resource use and environmental impact throughout the lifecycle of high-risk AI systems, including energy consumption. Article 11 in conjunction with Annex IV 3(b) AI Act EP Version mandates the disclosure of energy consumption

---

¹²³ <https://data.consilium.europa.eu/doc/document/ST-14954-2022-INIT/en/pdf> (last visited December 21, 2023)
¹²⁴ Nordhaus, "The climate club: how to fix a failing global effort", 99 Foreign Aff. (2020), pp. 12-13.
¹²⁵ Draft Compromise Amendments on the Draft Report Proposal for a regulation of the European Parliament and of the Council on harmonised rules on Artificial Intelligence (Artificial Intelligence Act) and amending certain Union Legislative Acts (COM(2021)0206 – C9 0146/2021 – 2021/0106(COD))
¹²⁶ Irish DPC, Decision of 31 December 2022, DPC Inquiry Reference: IN-18-5-7, Meta Platforms Ireland Limited (formerly Facebook Ireland Limited) in respect of the Instagram Service, 85 (violation of the data protection principle of fairness, Art. 5(1)(a) GDPR, as directed by the binding decision of the EDPB); Irish DPC, Decision of 31 December 2022, DPC Inquiry Reference: IN-18-5-5, Meta Platforms Ireland Limited (formerly Facebook Ireland Limited) in respect of the Facebook Service, 77-81 (principles of transparency and fairness, Art. 5(1)(a) GDPR).
¹²⁷ Cf., Bienert et al., Large AI Models for Germany: Feasibility Study https://leam.ai/feasibility-study-leam-2023/ (last visited December 21, 2023)

information during development and use, considering relevant Union and national legislation. The Trilogue Version of the AI Act introduced an additional requirement for general-purpose AI systems with systemic risk to document and report on the energy consumption of those models.[128] Importantly, and rightly, the Commission is charged to develop a methodology for calculating Key Performance Indicators and references for the Sustainable Development Goals (SDGs), including environmental impact (Recital 46b; see also Recital 87a AI Act EP Version). Harmonised standards according to Article 40 will be essential to effectively compare and evaluate the environmental impact of different AI systems.[129]

While transparency mechanisms such as logging and disclosing the greenhouse gas (GHG) footprint are a step in the right direction, numerous studies indicate that standard disclosures are often ignored by intended recipients.[130] Nevertheless, such mechanisms can be beneficial for non-governmental organisations, information intermediaries, and regulatory agencies seeking to collect data on the environmental impact of AI systems.

The fourth pillar encompasses risk assessment, specifically addressed in Article 28b(2)(a) AI Act EP Version. Providers of foundation models need to assess mitigate, and ultimately manage throughout the lifecycle (Article 28b(2)(f) AI Act EP Version), foreseeable risks not only with respect to health, safety, fundamental rights, democracy and the rule of law, but also the environment. While, as I have spelled out in detail elsewhere,[131] risk assessment and management should generally relate to specific use cases, addressing sustainability risks at the level of the model itself seems reasonable indeed: it is here that most resources for equipment and training are used, for example, for building a foundation model like GPT-4. In addition, any sustainability gains achieved early on in the model development may have a large impact that propagates down the AI value chain to all applications built on that specific (foundation) model. However, such a provision needs to be suitably operationalised. The proposal made below concerning sustainability by design and sustainable impact assessments (see Part IV.2.) precisely ties into and further develops the risk management framework proposed by the EP.

The fifth pillar involves guidance and review. Article 82b(1)(viii) AI Act EP Version calls for guidance by the European Commission on the practical implementation of environmental impact measurement and reporting methods, including carbon footprint and energy efficiency. This links to the standardisation efforts just mentioned. Article 84(3)(bf), in turn, pertains to review and potential update requirements. The Commission is tasked with including information on updated sustainability requirements in its biannual reports.

Notably, there is a missing pillar in the EU AI Act: operationalizing sustainability goals and translating them into effective action. While the Act addresses various aspects, it fails to provide a comprehensive framework that actively promotes meaningful action to address the climate change impact of AI. Thus, in my view, the proposed AI Act EP Version, and likely the final version of the AI Act based on the results of the Trilogue, still do not adequately address the environmental consequences of AI.

---

[128] Bertuzzi, AI Act: EU policymakers nail down rules on AI models, butt heads on law enforcement (EURACTIV, Dec. 7, 2023), https://www.euractiv.com/section/artificial-intelligence/news/ai-act-eu-policymakers-nail-down-rules-on-ai-models-butt-heads-on-law-enforcement/.

[129] Cf. Veale and Zuiderveen Borgesius, "Demystifying the Draft EU Artificial Intelligence Act—Analysing the good, the bad, and the unclear elements of the proposed approach" *Computer Law Review International 22 (2021)*, 97.

[130] See only Ben-Shahar and Schneider, More than you wanted to know. The Failure of Mandated Disclosure (Princeton University Press, 2014); Hacker, Verhaltensökonomik und Normativität: die Grenzen des Informationsmodells im Privatrecht und seine Alternativen (Mohr Siebeck, 2017), pp. 116, 132 et seqq.

[131] Hacker, Engel and Mauer, 'Regulating ChatGPT and other Large Generative AI Models' (2023) ACM Conference on Fairness, Accountability, and Transparency (FAccT '23) 1112, 1115 et seqq.

## IV. Policy proposals: sustainable AI regulation going forward

The preceding analysis has shown that current EU law, despite the applicability of technology-neutral environmental law and some promising steps in the latest version of the AI Act endorsed by the European Parliament, does not adequately address the climate risks posed by AI systems. This holds particularly true for GHG emissions. In the following section, this article introduces and discusses four policy proposals, ranging from co-regulation instruments, sustainability by design and restrictions on AI training to consumption caps based on an emissions trading regime.

### 1. Co-regulation

A first potential approach to regulating the use of AI and its impact on sustainability is through tools of co-regulation that have also been introduced in the GDPR. For example, Article 69 AI Act already encourages, as seen, the adoption of industry codes of conduct. These codes serve as voluntary agreements or standards developed and embraced by relevant stakeholders, such as companies, associations, or professional bodies, to guide their behaviour and practices regarding AI.[132] For instance, the GDPR encourages the creation of codes of conduct to promote sector-specific data protection rules (Art. 40 GDPR). Significantly, these codes may eventually be approved by national data protection authorities or even the Commission and thus be acquired general validity in the EU (Article 40(9) GDPR). Such a provision formalizing administrative oversight and endowing a limited binding effect on codes of conduct is missing in the AI Act and should be added to make them an attractive instrument, similar to a safe harbour provision.

The key advantage of industry codes of conduct is their ability to harness the distributed knowledge and expertise of the various actors involved in the design, development, and deployment of AI systems.[133] They can promote innovation and flexibility by allowing for customised solutions that are tailored to specific contexts and sectors, making them adaptable to evolving circumstances and needs.[134] Additionally, they can enhance trust and internalise the industry's commitment in addressing the ethical and social challenges posed by AI.[135]

However, industry codes of conduct also have their drawbacks. One of these is the potential lack of sufficient incentives for compliance, particularly when effective monitoring and enforcement mechanisms are absent.[136] Additionally, there is a risk of regulatory capture or fragmentation, as different groups or regions may adopt diverging or conflicting standards, which could undermine the coherence and consistency of the regulatory framework. Furthermore, these codes may not adequately represent the interests and values of all affected parties, such as consumers, workers, or civil society organisations, who may have limited participation or representation in the code development process.[137]

---

[132] Hagendorff, "The ethics of AI ethics: An evaluation of guidelines'" *Minds and Machines* 30 (2020), 99.
[133] Bartle and Vass, *Self-regulation and the regulatory state: A survey of policy and practice* (Centre for the Study of Regulated Industries, University of Bath School of Management, Research Report 17 2005), 48; Coglianese and Mendelson, 'Meta-regulation and self-regulation' in Robert Baldwin et al. (Eds.), *The Oxford Handbook of Regulation* (Oxford University Press, 2010), p. 164.
[134] Coglianese and Lazer, 'Management-based regulatory strategies' in John D. Donahue and Joseph S. Nye (Eds.), *Market-Based Governance* (Brookings Institution, 2002), p. 202.
[135] Ibid.
[136] Snyder Bennear, 'Evaluating management-based regulation: A valuable tool in the regulatory toolbox?' in Coglianese and Nash (Eds.), *Leveraging the private sector* (Routledge, 2010), p. 80.
[137] Glicken,"Getting stakeholder participation 'right': a discussion of participatory processes and possible pitfalls,"*Environmental Science & Policy* 3 (2000), 305.

Therefore, industry codes of conduct related to sustainable AI, and sustainability seals, should not be viewed as a substitute for binding regulation. Instead, they may complement compulsory measures and can enhance the effectiveness and legitimacy of AI governance. Going forward, it is essential that these codes may be endowed with general validity in certain sectors, offering sector-specific safe harbours; and that codes as well as seals undergo regular review and evaluation to ensure their ongoing relevance, reliability, and responsiveness to the evolving challenges and opportunities presented by the climate effect of AI.

### 2. Sustainability by design

A second and, in my view, even more promising proposal is to integrate a requirement of sustainability by design into the AI Act–either the version to be finalised until the beginning of 2024 or in the next update scheduled two years later (Article 84 AI Act). Inspired by the principle of data protection by design, sustainability by design aims to embed environmental considerations into the design and implementation of ML models and practices. As I will argue, a key tool for achieving this goal are sustainability impact assessments.

#### a) From data protection by design to sustainability by design

Over the past decades, data protection law has taken a compliance turn.[138] Data controllers not only need to answer to data subjects exercising their subjective rights, and agencies conducting inquiries, but have to establish technical and organisational routines to ensure data protection compliance even in the absence of individual and administrative proceedings (Art. 24 et seqq. GDPR). This is based on the correct assumption that subjective rights, and administrative inquiries, often come too late and are exercised too rarely to effectively protect data protection principles and data subject on the ground. One of the most notable provisions embodying this compliance term is the principle of data protection by design and default (Article 25 GDPR). Data protection, in this way, is converted from a mere reactive tool of ex-post control to a proactive instrument of ex-ante design.

As political scholars have pointed out repeatedly, however, civil liberties and freedoms, such as data protection, are essentially worthless if their grantees lack the material resources to exercise them and to flourish in the protective frame afforded by them.[139] This limitation resurfaces with renewed urgency in the current climate emergency. While the capabilitarian tradition rightly stresses access to basic amenities and resources, even these preconditions of the enjoyment of subjective rights are threatened for a growing number of persons by hostile environmental conditions as a result of climate change.

Hence, in my view, data protection by design needs to be complemented by "sustainability by design". At the technical and organisational level, developers need to ensure that all reasonable levers are pulled to minimise the contribution of ICT to climate change. Such a paradigm change has been explored for consumption practices[140] and product design[141] in the literature, and is increasingly translated into the practice of supply chain management and other industrial sectors for the pursuit of corporate ESG goals.[142] Building on these approaches, sustainability by design should also become a leading principle in the governance and regulation of the ICT sector. If we cannot solve the climate crisis, data protection by design will ultimately be a futile effort, a luxury game played out in a few privileged jurisdictions whose citizens–or courts and regulators–may still afford to care about data protection and privacy.

---

[138] See, e.g., Leta Jones and Kaminski, "An American's Guide to the GDPR" 98 Denv L Rev (2020), 93.
[139] See, e.g., Sen, *The Idea of Justice* (Harvard University Press, 2011).
[140] Ehrenfeld, Sustainability by design: A subversive strategy for transforming our consumer culture (Yale University Press, 2008).
[141] Vezzoli and Manzini, *Design for environmental sustainability* (Springer, 2008).
[142] See, e.g., <www.bcg.com/press/10february2022-bcg-cdp-build-tech-platform-scope-3-data-accelerate-decarbonization> (Last visited December 21, 2023); <www.accenture.com/us-en/services/sustainability/sustainability-by-design>, (last visited December 21, 2023); <www.designorate.com/principles-of-sustainable-design/> (last visited December 21, 2023).

### b) Sustainability impact assessments

As always with "by design" principles, the devil is in the details of implementing such ideals in concrete technologies and practices. In the context of AI regulation more specifically, mandatory sustainability impact assessments may be effective instruments to firmly integrate climate change considerations into the development of AI models.[143] Such a proposal may build on a large literature, and practical experience, concerning data protection,[144] social,[145] and algorithmic impact assessments.[146] While impact assessments are not a silver bullet[147] and embody important normative and design choices,[148] they do provide a promising route toward operationalizing sustainability considerations in the design and deployment of AI models.

More specifically, a mandatory sustainability impact assessment (SIA) should be a key component of the AI Act.[149] Indeed, the EP has added such wording for high-risk models (Art. 9(2)(a) AI Act EP Version) and for foundation models (Art. 28b(2)(a) AI Act EP Version). Both provisions call, inter alia, for a risk assessment and mitigation measures concerning environmental risks. While they present a step in the right direction, they should be simultaneously narrowed and expanded. First, Art. 28b(2)(a) AI Act EP Version should be restricted to an assessment of environmental hazards–GHG emissions and water consumption are the main risks which already materialise during the training of foundation models, before their actual deployment in real uses cases.[150] Any mistakes in this domain will inevitably propagate down the AI value chain. Other risks, e.g., to health, safety, and fundamental rights, are often best addressed at the concrete application level.[151]

However, the analysis of environmental risks more specifically should not be limited to high-risk (Art. 9(2)(a) AI Act EP Version) or foundation models (Art. 28b(2)(a) AI Act EP Version) only. Rather, the SIA should apply to developers of both high-risk and non-high-risk AI systems: the carbon footprint of AI systems is unrelated to their level of risk regarding health, safety, or fundamental rights, or their integration in Annexes II and III.

---

[143] See also Kaack and others, "Aligning artificial intelligence with climate change mitigation", *Nature Climate Change* 12 (2022), 518; Hacker, op. cit. *supra* note 4, 63 et seqq.
[144] See, e.g., Flaherty, "Privacy impact assessments: an essential tool for data protection" *Privacy Law & Policy Reporter* 5 (2000), 85; Wright and Hert (Eds.), *Privacy Impact Assessment* (Springer, 2012); Bieker and others, *A process for data protection impact assessment under the european General Data Protection Regulation* (Springer, 2016); Binns, "Data protection impact assessments: a meta-regulatory approach", *International Data Privacy Law* 7 (2017), 22; Van Dijk, Gellert and Rommetveit, "A risk to a right? Beyond data protection risk assessments" *Computer Law & Security Review* 32 (2016), 286; Gellert, "Understanding the notion of risk in the General Data Protection Regulation", *Computer Law & Security Review* 34 (2018), 279.
[145] See, e.g., Edwards, McAuley and Diver, "From privacy impact assessment to social impact assessment" 2016 IEEE (SPW) (2016), 53.
[146] See, e.g., Selbst, "An Institutional View of Algorithmic Impact Assessments", *Harvard Journal of Law & Technology* 35 (1) (2021), 117; Reisman and others, 'Algorithmic Impact Assessments: A Practical Framework for Public Agency' *AI Now* (2018); Kaminski and Malgieri, "Algorithmic impact assessments under the GDPR: producing multi-layered explanations2, (2018) *International Data Privacy Law,* 19; Moss and others, *Assembling accountability: Algorithmic Impact Assessment for the public interest* (APO Report, 2021).
[147] De Hert, "A human rights perspective on privacy and data protection impact assessments" in David Wright and De Hert (Eds.), *Privacy Impact Assessment* (Springer, 2012).
[148] Laux, Wachter and Mittelstadt, "Three Pathways for Standardisation and Ethical Disclosure by Default under the European Union Artificial Intelligence Act" (2023) *Working Paper*, <https://ssrncom/abstract=4365079>.
[149] See, e.g., Kaack and others, "Aligning artificial intelligence with climate change mitigation", *Nature Climate Change* 12.6 (2022).
[150] Another key risk at this level is discrimination; hence, Art. 10 AI Act should apply to foundation models as well. Similarly, content moderation should be addressed at the model level; see Hacker, Engel and Mauer, "Regulating ChatGPT and other Large Generative AI Models", Parts 6.3 and 6.4.
[151] Ibid, Part 7.2; but see also n. 150.

As part of the SIA, during the modelling phase, developers should compare different model and design types–such as linear regression versus deep learning,[152] federated versus non-federated learning,[153] or the use of a pre-trained models versus the training of a new model from scratch[154]–not only in terms of performance but also considering their estimated climate footprint.[155] As in the establishment of a design defect under product liability law,[156] only models and designs that are reasonably available to the developer, considering cost and utility, need to be integrated into the SIA. Similarly, downstream deployers should be under an obligation, when choosing between different models for a concrete application, to perform a sustainability impact assessment.

Obviously, such a constraint depends on the availability of comparing the GHG emissions of different models and design choices.[157] Fortunately, tools already exist to approximately measure the carbon impact of AI models.[158] Simply put, if two model types demonstrate similar performance, developers would be obliged under the new provision to choose the more sustainable model for further development and deployment. This approach would supplement the existing focus on performance measures with greater environmental awareness and practical, low-maintenance steps to integrate sustainability into the broader target function of ML development.

In fact, sustainability and performance may even align synergistically in many scenarios. As I have explained in detail elsewhere, the AI Act's rules on foundation models[159] and the concomitant liability provisions[160] need to be adapted to the specificities and complexities of foundation models, such as GPT-4. It should clearly focus on key risks to be addressed while developing foundation models: climate risks, including SIAs; discrimination; transparency; cybersecurity; and toxicity.[161] In this way, the most urgent societal costs of generative AI can be tackled, and performance-oriented AI development be complemented by trustworthiness *and* sustainability.

---

[152] Strubell, Ganesh and McCallum, "Energy and policy considerations for modern deep learning research", *Proceedings of the AAAI conference on artificial intelligence* 34(09) 2020, pp. 13693-13696.

[153] Güler and Yener, "A framework for sustainable federated learning" (IEEE, 2021).

[154] Cf. Patterson and others, "Carbon emissions and large neural network training" (2021) arXiv preprint arXiv:210410350.

[155] The exact impact is not easy to measure. An index including Scope 1, 2 and 3 Emissions for necessary compute resources (e.g., energy; carbon emissions) for training and retraining could be used as a proxy. For a more comprehensive impact measure (including production, transport, and end-of-life, as well as water consumption), see OECD, op. cit. *supra* note 16, Annex A; on Scope 1, 2 and 3 Emissions, see IPCC, Working Group III Contribution to the Fifth Assessment Report of the Intergovernmental Panel on Climate Change (2014), 122.

[156] Hacker, op. cit. *supra* note 4, 24 et seq.

[157] Water consumption should be monitored as well, to the extent possible. Any trade-offs between GHG emissions and water consumption (see, e.g., Li and others, op. cit. *supra* note 6, 9; below, text accompanying n. 163) have to be resolved on a case-by-case basis, grounded in best practices, a reasonable weighting of all relevant factors, and regulatory guidelines.

[158] Overview in OECD, op. cit. *supra* note 16, 28; Kasper Groes Albin Ludvigsen, 'How to estimate and reduce the carbon footprint of machine learning models' Towardsdatascience <https://towardsdatascience.com/how-to-estimate-and-reduce-the-carbon-footprint-of-machine-learning-models-49f24510880>, (last visited 30 August 2023); popular tools include https://codecarbon.io/ and https://mlco2.github.io/impact/.

[159] [anonymized]

[160] [anonymized]

[161] [anonymized]

### 3. Restrictions on training AI models

One further entry point for a regulation is the time, location and type of training large AI models undergo. As mentioned, the large number of iterations necessary to calibrate state-of-the-art machine learning models consumes significant amounts of energy and freshwater–leaving a large GHG footprint. Hence, GHG emissions of AI training might be lowered by (requiring) a shift to regions where large amounts of renewable are available to power AI training.[162] In this vein, one may envision a provision binding AI developers might to conduct training only at facilities that derive a certain percentage of their energy from renewable sources.

The devil, however, is again in the details. One particularly attractive move to raise the renewable energy percentage would be to "follow the sun", i.e., to conduct training in regions with excellent opportunities for photovoltaic production of energy. Incidentally, these regions may also, precisely because of their sun exposure, exhibit higher average temperatures–leading to greater water needs to cool data centres.[163] Hence, factoring water consumption into the equation introduces potentially hard trade-offs between the conservation of scarce sources and GHG emissions. Federated learning strategies may offer a way out of this impasse, as they can be both water-[164] and energy efficient.[165] Ultimately, these questions will probably have to be settled in a country- and region-specific way, depending on the available resources. For the time being, however, rules such as a minimum threshold of, e.g., 50% of renewable energy powering AI training should be considered, and federated learning strategies should be explored and incentivised further. Evidently, such minimal thresholds could and should be simultaneously introduced for other technologies and industries, depending on the availability of renewable energies in specific regions.

### 4. Consumption caps

The final, most intrusive, but potentially also most effective regulatory mechanism contemplated here to render AI more sustainable is the establishment of energy consumption caps. This holds particularly as the establishment of an EU wide carbon tax–the second theoretically and empirically proven mechanism for cost-effective GHG reduction through carbon pricing[166]–remains politically elusive.[167]

#### a) An emissions trading regime for AI

The most straightforward way to incentivise limits on GHG emissions is to include AI, and potentially ICT more generally, in the EU ETS. In fact, such an inclusion seems justified. The ETS currently applies to commercial aviation–but as seen, ICT emissions are likely on par with or above those of commercial aviation, and they are likely to rise more steeply. For all the criticism rightly levied against its imperfections,[168] the ETS still represents the easiest and most feasible mechanism to systematically price GHG externalities and to set workable financial incentives to reduce them.[169] Significantly, lessons must be learned from the mixed trajectory and results of the current EU ETS. For example, initial caps need to be set low enough to encourage trading, raise allowances prices, and incentivise GHG savings.[170] A carbon price floor can partially remedy market failures.[171] While the details (e.g., defining the exact type of ICT covered) deserve further scrutiny and transcend the scope of this paper, the rising importance of AI and ICT-related GHG emissions speaks in favour of including them in the ETS rather sooner than later.

---

[162] Li and others, op. cit. *supra* note 6, 9.
[163] Ibid, 9.
[164] Ibid, 8.
[165] Güler and Yener, "Sustainable federated learning" (2021) arXiv preprint arXiv:210211274.
[166] See, e.g., Haites, "Carbon taxes and greenhouse gas emissions trading systems: what have we learned?", 18 Climate Policy (2018), 955, 955 et seq.
[167] See, e.g., Skou Andersen, "The politics of carbon taxation: how varieties of policy style matter", 28 Environmental Politics (2019), 1084; Bell and others, *Environmental Law* (OUP, 2017)*,* p. 550.
[168] See only Böhringer, "Two decades of European climate policy: A critical appraisal", 8 Review of Environmental Economics and Policy (2014), 1; Robert Baldwin, "Regulation lite: The rise of emissions trading", 2 Law and Financial Markets Review (2008), 262, 264 et seqq.

### b) Caps based on an AI model's social utility

A more challenging approach would consist in differentiating between different AI-based use cases. ML systems have become ubiquitous in various domains, such as healthcare, education, entertainment, and security. Arguably, these sectors differ in their criticality for basic societal tasks. Hence, consumption caps could be formulated as a function of the sector and specific use case the model is deployed in–similar to the risk qualification undertaken in Art. 6 and Annexes II and III AI Act. Significantly, many of the areas listed in Annexes II and III are, arguably, among the most important societal sectors in which machine learning could, if properly used, simultaneously have the greatest positive societal impact, e.g., medicine, employment, administration, transportation and automotive, etc.

As a first step, lawmakers or regulators would have to define certain "social usefulness classes" based on the expected societal benefits of harnessing AI in a certain area. This approach raises the question of how to measure and compare the social value of different ML applications. Ultimately, this is a judgment call that depends on what we can and wish to afford as a society in terms of energy and carbon emissions. It is a debate that, as a consequence of the climate emergency, our societies must, however, increasingly be prepared to have.

While any such classification will be contentious, at least three broad classes–high-benefit; medium-range; low-benefit–could be established fairly rapidly. Borderline cases can never be avoided, but in many scenarios, an allocation to these three main buckets should be fairly uncontroversial. While most of the use cases listed in Annexes II and III AI Act should belong to the high-benefit category, use cases related to entertainment, marketing and advertising can be generally relegated to be low-benefit bracket. Most other cases would then populate the medium-range category.

In a second step, soft or hard consumption caps could be allocated to these classes. They would designate the amount of energy that can be used, or GHG that may be emitted, to train and run an ML system in that specific sector and use case, similar to the emissions permits regime established by the Industrial Emissions Directive 2010/75/EU.[172] In this way, the permitted climate costs of an AI model would depend on how valuable and beneficial the system is for society.

The proposal is complicated further because, particularly in AI value chains where foundation models are fine-tuned, emissions cannot be simply mapped onto a specific use case as the climate costs of the foundation model factors into potentially a great variety of different applications. However, such measuring problems can, in my view, be overcome. One may either use a specific fraction (e.g., 1/1000) of the foundation models costs for each use case, or attempt to specifically measure only the cost for fine-tuning, deploying and running the model for a specific use case, with correspondingly lower consumption caps.

---

[169] See, e.g., Perman, *Natural Resource and Environmental Economics* (Pearson Education, 2003), pp. 333 et seq.; Gawel, "Der EU-Emissionshandel vor der vierten Handelsperiode–Stand und Perspektiven aus ökonomischer Sicht", 8 Zeitschrift für das gesamte Recht der Energiewirtschaft (2016) 351, 357.
[170] ibid, 352.
[171] Hintermayer, "A carbon price floor in the reformed EU ETS: Design matters!", 147 Energy Policy (2020), Article 111905; Parry, Black and Roaf, "Proposal for an international carbon price floor among large emitters" (IMF Staff Climate Note No 2021/001, 2021).
[172] See, e.g., Art. 4 of the Directive 2010/75/EU of the European Parliament and of the Council of 24 November 2010 on industrial emissions (integrated pollution prevention and control) (recast), O.J. 2020, L 334/17; Bell and others, op. cit. *supra* 186, 495 et seqq.

Eventually, I submit, there should indeed be a difference in climate costs permissiveness between training an ML system for medical diagnostics that could potentially, with the right guardrails in place,[173] save hundreds of lives, and developing an ML system for mere entertainment purposes. For example, this difference could also translate into greater leeway in the sustainability impact assessment that would be required before deploying an ML system. By giving more flexibility to ML systems with higher social utility, we could encourage more innovation and research in areas that are critical for human well-being and social welfare, while balancing the expected benefits of specific AI applications with their climate costs.

### 5. A mix of instruments and interpretations: the interplay between solutions

No unique silver bullet exists to fully address the energy and water consumption of large AI models. Hence, the different instruments sketched here will have to be combined in a comprehensive strategy to mitigate the environmental impact of AI while capturing its potential for environmentally beneficial use cases. For example, co-regulation (codes of conduct) can be used to specifically implement sector-specific thresholds and metrics for sustainability-by-design approaches, and to operationalize any obligations the AI Act might bring along. Similarly, energy consumption caps may be raised if more energy is derived from renewable sources.

How do these measures, then, relate to the environmentally-aware interpretation of the GDPR in this context? First, as long as the policy proposals have not been put into place, the suggested interpretation of the GDPR may do some of the heavy lifting in specific areas, such as model retraining. Second, even once a more comprehensive framework has been enacted, the environmentally-aware interpretation of the GDPR is, arguably, still necessary because no other provisions allow for a trade-off between individual data subject rights and collective interests.

## V. From sustainable AI to sustainable technology regulation

AI models are, however, not the only digital tools or infrastructures entailing significant GHG emissions; hence, sustainable AI regulation may provide a framework for other technologies.

### 1. Sustainability challenges in digital technology

Indeed, many other digital technologies face significant sustainability challenges due to their high GHG emissions: for example, blockchain (e.g., Bitcoin);[174] metaverse applications;[175] and data centres (irrespective of the technology they serve),[176] to name only a few. Already in 2005, the energy demand of the global force of data centres equalled 1% of the entire energy consumption of the US and the GHG emissions of Argentina.[177] In 2010, they were estimated to use between 1.1% and 1.5% of total global electricity.[178] Moreover, their consumption is expected to grow substantially over the next years as cloud services and large compute infrastructures become ever more important to sustain advanced, data-intense IT applications, such as large AI models or the metaverse.[179]

---

[173] See, e.g., Mittelstadt, Wachter and Russell, "The Unfairness of Fair Machine Learning: Levelling down and strict egalitarianism by default", (2023) arXiv preprint arXiv:230202404.
[174] Sedlmeir and others, "The energy consumption of blockchain technology: Beyond myth", 62 Business & Information Systems Engineering (2020), 599; Truby, "Decarbonizing Bitcoin: Law and policy choices for reducing the energy consumption of Blockchain technologies and digital currencies", 44 Energy Research & Social Science (2018), 399.
[175] Bertuzzi, 'EU Commission moves to link 'fair share' debate with the metaverse' (2023) <www.euractiv.com/section/digital/news/eu-commission-moves-to-link-fair-share-debate-with-the-metaverse> (last visited 10 Jan. 2023).
[176] Technology Policy Council, *TechBrief: Computing and Climate Change* (Association for Computing Machinery, 2021), p. 1.

### 2. Sustainable AI regulation as a blueprint

In these areas, sustainable AI regulation may arguably serve as a blueprint. The regulatory toolbox described above, from co-regulation to sustainability assessments, functioning requirements and hard consumption caps, can and should be flexibly adapted to these other areas of technology law.

For example, emissions trading famously only applies to certain sectors. With digital technology permeating all sectors, and core infrastructures like data centres serving them all, it seems advisable to extend the scope of application. Preferably, it should encompass ICT device production, processes, and infrastructures in general. To the very least, high-consuming structures–like data centres and blockchain mining centres–should be covered to set effective economic incentives to lower GHG emissions. In the case of crypto applications primarily used for speculation and money laundering (e.g., certain NFTs),[180] even quite strict hard consumptions caps should be considered. Overall, the transversal structure and effects of global ICT require new solutions with a view to their rising contributions to climate change–preferably solutions on a global scale. But the EU should at least make a first move in the absence of effective global coordination, mechanisms, and frameworks.

## VI. Conclusion

This paper suggests that AI regulation should complement its core goal of trustworthiness with an ambitious sustainability objective. This desideratum has acquired urgency with the advent of large generative AI models like ChatGPT or GPT-4. While such large AI models may offer sustainability benefits in the long run, their training and deployment is highly resource intensive along several parameters, such as GHG emissions and water consumption.

Against this background, the paper articulates a framework for sustainable AI and technology regulation. While the key instruments of extant environmental law do not explicitly cover AI and emerging technologies, those technology-neutral obligations partially apply to GHG emissions and water consumption of advanced digital technology, but could and should be adapted more specifically to it. Using AI development and deployment as a lead example, I show that the significant challenges AI poses regarding environmental sustainability can, to a certain extent, be taken into account by a climate-aware interpretation of the GDPR and non-discrimination law. Beyond this, they are only inadequately addressed by the upcoming EU AI Act. Hence, the paper suggests and develops a range of strategies, from co-regulation, to sustainability by design, including sustainability impact assessments, restrictions on AI training and consumption caps.

---

[177] Mathew, Sitaraman and Shenoy, "Energy-aware load balancing in content delivery networks", (2012) 2012 Proceedings IEEE INFOCOM 954, 954.

[178] Corcoran and Andrae, "Emerging trends in electricity consumption for consumer ICT", (2013) National University of Ireland, Galway, Connacht, Ireland, Tech Rep, 7.

[179] Dayarathna, Wen and Fan, "Data center energy consumption modeling: A survey", 18 IEEE Communications Surveys & Tutorials (2015) ,732, 732.

[180] Dupuis and Gleason, "Money laundering with cryptocurrency: open doors and the regulatory dialectic", 28 Journal of Financial Crime 60 (2020); Jordanoska, "The exciting world of NFTs: a consideration of regulatory and financial crime risks", 10 Butterworths Journal of International Banking and Financial Law (2021) 716.

Within sustainability by design strategies, one important mechanism could be what I term "sustainability impact assessments". Crucially, during the modelling phase, developers should compare different AI model types (e.g., linear regression versus neural networks) not only regarding their performance but also their estimated GHG footprint. Consumption caps, on the other hand, may be implemented by expanding the reach of the EU's emissions trading system to AI as well as other digital technology and infrastructure, such as data centres, metaverse applications, or blockchain. Such a comprehensive framework is needed to simultaneously tackle and coordinate the dual fundamental societal transformations of our time: digitisation and climate change mitigation.